\RequirePackage{pdf14}
\documentclass[acmsmall]{acmart}
\usepackage{amsfonts}
\usepackage{colortbl}
\usepackage[utf8]{inputenc}
\usepackage{array,multirow,colortbl}
\usepackage{siunitx}
\usepackage[export]{adjustbox}
\usepackage{amssymb}

\usepackage{tabularx}
\usepackage{graphicx}
\usepackage{xcolor}
\newcommand{\an}[1]{\textcolor{black}{#1}}
\newcommand{\rev}[1]{\textcolor{black}{#1}}

\usepackage{tikz} 
\usepackage{xcolor}
\usepackage[title]{appendix}

\newcommand{\mybox}[4]{
    \begin{figure}[H]
        \centering
    \begin{tikzpicture}
        \node[anchor=text,text width=.9\columnwidth, draw, line width=1pt, fill=#3, inner sep=4mm] (big) {\\\small#4};
        \node[draw, line width=.5pt, fill=#2, anchor=west, xshift=5mm] (small) at (big.north west) {#1};
    \end{tikzpicture}
    \end{figure}
}

\def\BibTeX{{\rm B\kern-.05em{\sc i\kern-.025em b}\kern-.08emT\kern-.1667em\lower.7ex\hbox{E}\kern-.125emX}}
    
\setcopyright{acmlicensed}
\acmJournal{PACMHCI}
\acmYear{2019} \acmVolume{3} \acmNumber{CSCW} \acmArticle{74} \acmMonth{11} \acmPrice{15.00}\acmDOI{10.1145/3359176}
\received{April 2019} 
\received[revised]{June 2019}
\received[accepted]{August 2019}

\begin{document}

%
\title{Collaboration Drives Individual Productivity}

%
\author{Goran Murić}
\affiliation{%
  \institution{USC Information Sciences Institute, USA}
  \city{Marina Del Rey}
  \state{California}
}
\email{gmuric@isi.edu}

\author{Andrés Abeliuk}
\affiliation{%
  \institution{USC Information Sciences Institute, USA}
  \city{Marina Del Rey}
  \state{California}
}
\email{aabeliuk@isi.edu}

\author{Kristina Lerman}
\affiliation{%
  \institution{USC Information Sciences Institute, USA}
  \city{Marina Del Rey}
  \state{California}
}
\email{lerman@isi.edu}

\author{Emilio Ferrara}
\affiliation{%
  \institution{USC Information Sciences Institute, USA}
  \city{Marina Del Rey}
  \state{California}
}
\email{ferrarae@isi.edu}

%
\renewcommand{\shortauthors}{Goran Murić et al.}

%
\begin{abstract}
How does the number of collaborators affect individual productivity? Results of prior research have been conflicting, with some studies reporting an increase in individual productivity as the number of collaborators grows, while other studies showing that the {free-rider effect} skews the effort invested by individuals, making larger groups less productive. The difference between these schools of thought is substantial: if a super-scaling effect exists, as suggested by former studies, then as groups grow, their productivity will increase even faster than their size, super-linearly improving their efficiency. We address this question by studying two planetary-scale collaborative systems: GitHub and Wikipedia. By analyzing the activity of over 2 million users on these platforms, we discover that the interplay between group size and productivity exhibits complex, previously-unobserved dynamics: the productivity of smaller groups scales super-linearly with group size, but saturates at larger sizes. This effect is not an artifact of the heterogeneity of productivity: the relation between group size and productivity holds at the individual level. People tend to do more when collaborating with more people. We propose a generative model of individual productivity that captures the non-linearity in collaboration effort. The proposed model is able to explain and predict group work dynamics in GitHub and Wikipedia by capturing their maximally informative behavioral features, and it paves the way for a principled, data-driven science of collaboration.
\end{abstract}

%
%


%

%

%
\maketitle

\section{Introduction}

The benefits of collaboration and teamwork have been well documented. Groups can benefit from the diversity of their members~\cite{page2007power}, while the collective intelligence effects~\cite{Woolley2010} enable a group to outperform even its most capable individual members~\cite{Laughlin2006, Carey2012}. A group is better able to meet the multidisciplinary demands of today's increasingly more complex scientific problems, as seen in the growing shift to larger groups in science~\cite{Wuchty2007}. \an{Open-source collaboration platforms exhibit a shift in the distribution of work, with a few ``elite'' users doing most of the work
~\cite{kittur2007power}.} There is additional evidence of the fundamental differences in how groups operate within their respective fields based by their size: small groups produce more disruptive, innovative and potentially risky work, while larger groups tend to build on the  existing concepts~\cite{Wu2019}.

Although groups have an advantage over individuals, it is unclear whether larger groups have an advantage over smaller groups. Even within the domain of software development, where teams of developers can vary in size up to hundreds of people, there are no clear answers to this question. Smaller teams are more agile and require less communication overhead to coordinate~\cite{Faraj2000}. Smaller teams also avoid the {``free-rider''} effect~\cite{Strong1990}, which leads  individual  members of larger teams to lower their effort if they perceive no impact (or proportionally lower impact) of their contributions~\cite{Shepperd1993}. In contrast, other research has shown that team size has a positive effect on the total amount of work invested in software developments projects~\cite{Klug2016, Pendharkar2007, Sornette2014}. Understanding the relationship between group size and  performance can help improve productivity yielding policies to create optimal groups, e.g., by analyzing the cost-benefit of adding new collaborators versus upgrading the information technology~\cite{tohidi2006productivity}.

In a quest to disentangle the relationships between group productivity and its confounding factors, some researchers have suggested that the group's effort on a software project scales super-linearly with its size~\cite{Pendharkar2008,Sornette2014}. This is an important finding, as it suggests that doubling a group's size would more than double the amount of work it produces. In contrast, other studies found that productivity of software developers decreases with increasing group size~\cite{Scholtes2016,maxwell1996software}, and argue this is due to communication and coordination costs. This disagreement may be partly rooted by differences on the productivity measures, the unit of time analyzed, and the definition of a group~\cite{maillart2016aristotle}. More importantly, these previous studies only analyzed a few hundred software development groups working on highly popular software projects. In this paper, we address these issues by expanding the analysis to millions of groups across two open-source collaboration platforms.

\mybox{\textbf{Significance of our results}}{gray!30}{gray!10}{Collaboration is one of the foundations of social behavior. Understanding how the performance of groups changes as they grow is key to optimizing productivity of organizations at all scales. Our analysis of the planetary-scale GitHub and Wikipedia collaboration platforms reveals complex super-scaling dynamics for small groups, and marginal returns on productivity of increasingly larger groups. We closely examine the marginal gains of the individual efforts, and report the unexplored non-linear behaviors.}

\subsection*{Collaboration and teamwork}
Broadly speaking, a team can be defined as a group of individuals who coordinate to perform activities to achieve some common goals. The teams can differ in many ways; however, we can generally place them on a spectrum where on one side there are \textit{structured} and on the other side there are \textit{unstructured} teams. We can define the structured teams to have precisely defined roles, organized in advance by an external entity, with predefined communication patterns and management hierarchy. Such highly structured teams are found in traditionally organized businesses or in the military. In contrast, unstructured teams have no predefined roles and no management structure. The collaboration groups we focus on are usually somewhere in between. The emergence of the large-scale collaboration platforms, such as GitHub or Wikipedia, gave rise to a new type of \textit{loosely structured} teams. Such teams, or collaboration groups, differ from ``traditional'' teams in several respects: explicit communication between collaborators may not exist; coordination is not established in advance, but rather ad-hoc and organically; collaborators often have no predefined roles; motivations for contribution are often intrinsic. 
Therefore, we refer to such units as \textit{groups of collaborators} rather than teams, even though they do exhibit some characteristics of structured teams.

\section{Results}
Motivated by open questions and the availability of large-scale collaboration data, we address the problem of quantifying the relationship between number of collaborators and group productivity in open-source collaboration environments. We study two planetary-scale online platforms: \textit{GitHub}, a collaboration platform, where individuals work on solo software projects or collaborate in teams, and \textit{Wikipedia}, the World's largest encyclopedia written collaboratively by a large community of contributors. By analyzing the activity of approximately two million users and three million projects on GitHub and more than 700 thousand users collaborating on 2.6 million pages on Wikipedia, we discover that the interplay between number of collaborators and productivity exhibits complex and nonlinear dynamics, yet, both platforms display distinct commonalities.   Having more collaborators is usually beneficial. We show that the marginal benefit from each additional collaborator is \textit{super-linear}. Such benefit, however, declines as the group grows. We show that the productivity of a group as a whole is the result of increased productivity of \textit{all individual members}, and not simply due to the presence of a few super-performing individuals. The relationship between size and productivity is not an artifact of averaging over heterogeneous group members, but  holds at the individual level. When given the choice to work in multiple groups, users put more effort into projects with more collaborators. Users increase their individual activity with increasing group size, and each member is significantly more productive in a large group than when working alone or in a small group. Additionally, we analyze robustness of this finding by looking at a variety of factors which affect user's productivity.

\section{Related Work}
Laughlin and Carey~\cite{Laughlin2006, Carey2012} demonstrate that groups almost always perform better than the best individuals when solving an analytical task, both in performance and speed. Groups also tend to outperform a set of an equivalent number of independent workers~\cite{Mao2016}. Even though each member of a group puts less effort in on average, the gains of a collaboration usually dominate such losses.

For a number of tasks, besides the necessary skills, creativity plays an important role. In that case, in addition to the group size, other factors have to be considered, such as experience and time spent in a group. Creativity gets stimulated when new innovators join the group, bringing the new solutions for some old problems. However, one should keep in mind the balance of diversity in a group, as highly diversified groups could be prone to miscommunication. Besides all of the challenges, even in creative environments such as science and art, groups perform better~\cite{Guimera2005}. Wuchty et al. \cite{Wuchty2007} show that teams in science produce the exceptionally high-impact results. Teams also typically produce more frequently cited research than individuals do.

In general, groups have an advantage over individuals. However, it is still unclear if larger groups have an advantage over smaller groups. The answer to that question is complex and domain specific. Larger groups are more robust against external challenges~\cite{Derex2013}. Larger groups are also able to gain more success in collaborative efforts such as software development~\cite{Klug2016}. On the other hand, smaller groups are more agile which makes them more adaptable and therefore suitable for various ad-hoc projects. The benefits of small, flexible groups are particularly evident in the case of human conflicts~\cite{Bohorquez2009}. In the area of software development, team sizes can vary from a couple of developers to hundreds of them. Having a larger team seems to have a positive impact on the project success~\cite{Klug2016}, but at the same time increasing a team requires effective team coordination~\cite{Faraj2000}. 
\an{For example, in Wikipedia, having more editors work on an article was associated with an increases in article quality only if the planning is done by a small subset of the contributors~\cite{kittur2008harnessing}.}
Both software size and team size independently have increasing returns to scale relationship with software effort~\cite{Pendharkar2008}.

Even for a single professional domain, the size of a group can affect the collective productivity in non-linear, multi-directional and conflicting ways. Increasing the group size usually has a positive effect on a total amount of work invested by the collaborators~\cite{Klug2016,Pendharkar2007,Sornette2014}. The underlying stimulants which drive such a trend can be explained as the motivation of an individual when collaborating with others~\cite{Hertel2000}. On the other hand, group size can affect the individual and therefore total performance in the opposite way. The productivity loss occurs when individuals perceive no value to contributing or perceive no impact of their contributions to achieve the desired outcome~\cite{Shepperd1993}. The members of a large community tend to invest less effort in average due to the effect of \textit{social loafing}~\cite{Shiue2010}. Known also as \textit{free-riding}~\cite{Strong1990} or \textit{lurking}~\cite{oliveira2018exchange}, it affects the communities where the effort of a fraction of individuals compensate for the rest who do not contribute. Furthermore, the late addition of new members often reflects negatively on the productivity~\cite{Brooks} \an{and are more likely to drop out due to social barriers~\cite{steinmacher2019overcoming}.} A decrease in the productivity can be attributed to communication overhead~\cite{DiPenta2007}, which becomes prominent in larger groups. If the communication is unstructured and loosely organized, the amount of information each member of a group has to process can lead to the information overload~\cite{Nematzadeh2016}, which trumps the quality of communication and subsequently, the collaboration can be affected.

\section{Data and notation}

\subsection{GitHub}
GitHub is a Web-based system for version control. It allows multiple users to work on joint projects while keeping track of the full project history and versions. GitHub allows the collaborative work of teams of various sizes on a number of project types. It is mainly oriented towards the developer community and used for facilitating the development of software, both open source and proprietary, as it supports distributed non-linear workflows. Projects on GitHub are called \textit{repositories} and they contain the set of files related to the particular project. Besides the collaborative functions, GitHub provides a communication platform which allows developers to report issues or comment on other repositories. Users of GitHub can initiate various actions on repositories, such as \textit{pull}, \textit{push}, \textit{commit}, \textit{watch}, etc. In our analysis, we focus on users who perform \textit{push} action. A \textit{push} is a group of code updates uploaded to the repository. It can be performed only by users who are granted a privilege to make changes on the code. Those are the users who are trusted by the owner of the repository and make a part of the team. In our analysis, we focus only on those users considered to be the \textit{active collaborators}. We define active collaborators as users who have an autonomy to modify the repository directly by performing a \textit{push} action. \an{Therefore, we did not include \textit{pull requests} as they can be initiated by any member of a community and they are relatively rare~\cite{Kalliamvakou2014}.} Notice that each \textit{push} can contain an unspecified amount of code updates. However, we use it as a proxy to identify the group of developers working together. \an{The lines of code and the number of commits are some of the alternative measures used in literature. Previous research has shown that any of those units could be used to measure work, as the majority of pushes consist of a single commit~\cite{Klug2016} and the relation between lines of code and commits exhibits the same scaling~\cite{Sornette2014}.}

\textbf{GitHub Dataset.} We used a dataset consisting of all public actions performed on GitHub in the period of 32 months starting from January 1st 2015, recorded on 1s intervals. The data used for the analysis is made of all push actions performed by users on a subset of repositories during the first three months after the repository creation. The dataset consists of \texttildelow 40.9 million records of \textit{push} actions performed by \texttildelow 2 million users on \texttildelow 4.7 million repositories. Choosing the optimal time span for analysis is a result of a trade-off between: \textit{temporal relevance}, \textit{contextual relevance}, and \textit{sample size}. \textit{Temporal relevance:} we use the most relevant data available as in average more than 90\% of all events are performed during the first three months (Figure~\ref{fig:fraction_contributions} in Appendix~\ref{sec:activity}). \textit{Contextual relevance:} we define a group of collaborators working on a project on GitHub as a set of users actively contributing to a project over a relatively short period of time. We consider very late contributions to be less relevant to the actual group dynamic. \textit{Sample size:} the period of three months allows the analysis of a significant number of groups of various sizes.  \rev{Selecting different time windows yields similar behavior with different scales. We perform a sensitivity analysis for varying window time in Appendix~\ref{sec:sensitivity_analysis}.}

Additionally, we use data on user profiles and repository profiles. User profiles contain supplementary information about the user, such as a date of profile creation, the total number of created repositories, etc. Repository profiles contain information such as a date of repository creation, number of watchers, programming languages used in a repository, etc.

\subsection{Wikipedia}
Wikipedia is the largest and most popular online encyclopedia. The encyclopedic articles, which are referred to as \textit{pages} are the result of collaborative effort of volunteers, who are allowed to edit entries. Any user can become an editor by visiting any page and click on \textit{``Edit''}. Initially, it started as a completely open system where all authors had the same privileges and anyone could edit any page. However, certain limitations were quickly introduced to prevent undesirable edits of some particularly controversial, sensitive and/or vandalism-prone pages~\cite{wiki}. Every page has a corresponding \textit{talk} page, where the editors can communicate between each other, usually by writing an additional notes about their edits. Besides the regular users, there is a number of bots which can edit pages. The bots on Wikipedia are the automated tools that carry out certain automated tasks, such as: fixing errors, updating data or reverting vandalism. In our analysis we focus on edits performed only by registered users, excluding edits performed by bots and similar scripts which have a user status, but are actually machines. \an{Besides the number of edits, an alternative unit is the size of the edit measured by the number of bytes changed. Additional analysis described in Appendix~\ref{sec:wiki_scaling} show that there is a similar scaling relation between number of edits and size of edits and that both of them could be used as a valid measure of contribution.} 

\textbf{Wikipedia Dataset.} The data used for this analysis is extracted from December 2018 dump of English Wikipedia edits. The dump contains all relevant revision metadata on more than 60 million pages from their creation to date. Besides the data on main articles (5.8M), the dump contains edit history of other page types such as: \textit{talks}, \textit{user pages}, \textit{user talks}\ldots

Here, we analyze a sample of main Wikipedia pages chosen uniformly at random. We analyze \texttildelow 2.6 million pages with \texttildelow 23 million edits performed by \texttildelow 700 thousand users. We exclude \textit{redirect} pages, as their purpose is to automatically send visitors to another page, usually an article or section of an article. In average, Wikipedia article is 105 months old and the time difference between first and last edit is 93 months. Such a longevity allows contributions from multiple users over the period of many years. One can argue that the editors who make the updates long after each other do not collaborate. Our focus is on collaboration where people work in structured or unstructured teams, meaning that they make edits successively or alternately in relatively short period of time. In average more than 36\% of all edits are performed in just three months after the page creation (Figure~\ref{fig:fraction_contributions} in Appendix~\ref{sec:activity}) and therefore we focus our analysis only in the first three months of edits, regardless of time of creation. \rev{Similarly to GitHub, our results hold for varying time windows. More details in Appendix~\ref{sec:sensitivity_analysis}.}

\subsection{Notation and Definitions}
Let us define some terms we use throughout this paper. A brief overview of all relevant definitions, for future reference, is summarized in Table~\ref{tab:notation}.

\begin{table}
\caption{A short overview of terms and notations used in the paper}
\footnotesize
\begin{tabularx}{\linewidth}{rX} 
\textbf{Term} & \textbf{Definition} \\ 
\midrule
\textit{User}, $u$ & An active member of GitHub or Wikipedia \\
\textit{Project}, $P$ & Either a repository on GitHub or a page in Wikipedia \\
\textit{Number of projects}, $P_s$ & Number of projects the user is contributing to \\
\textit{Group}, $G$ & A set of collaborators who work on the same project \\
\textit{Group size}, $N$ & The number of collaborators belonging to a group, $N=|G|$.\\
\textit{Work}, $w$ & Number of contributions made by a user. The work of a group is $W= \sum_{i} w_{i}$ \\
\textit{Effective group size}, $n$ & $n=2^{H}$, where $H= - \sum_{i=1}^{N}f_{i} \log_{2}f_{i}$, and $f_{i}=w_i/W$ \\
\textit{Aggregate focus}, $F$ & $F = \sum_{i=1}^{N}c_{i}$, and $c_{i}=w_{i}/\sum_{i,k} w_{i,k}$ \\
\textit{User age} , $A_u$ & Time in days measured from the creation of user account to the end of the data range \\
\textit{Watchers}, $W_c$  & Number of watchers of a repository on GitHub \\
\textit{Forks}, $F_o$ & Number of times a repository was "forked" \\
\textit{Repository age}, $A_r$ & Time in days measured from the creation of a repository on GitHub to the last recorded push within the data range \\
\textit{Followers}, $F_l$ & Number of followers of a user on GitHub \\
\textit{Owned repositories}, $O_r$ & Number of repositories on GitHub the user created \\
\textit{Repository description}, $D_{sc}$ & Number of characters in a GitHub repository's description field \\
\textit{Page age}, $A_p$ & Time in days measured from the creation of a page on Wikipedia to the last recorded edit within the data range \\
\textit{Largest group}, $N_{max}$ & The largest group of collaborators the user is working with\\
\textit{Smallest group}, $N_{min}$ & The smallest group of collaborators the user is working with\\
\textit{Edit size}, $E_s$ & The size of an edit on Wikipedia page measured in bytes\\
\textit{Number of created pages}, $P_c$ & The number of pages on Wikipedia a user created\\
\bottomrule
\end{tabularx}
\label{tab:notation}
\end{table}

\textit{User} ($u$) is an active user of GitHub or Wikipedia who performed at least one \textit{contribution} during the observed time period, and $u \in \mathbb{U}$, where $\mathbb{U}$ is a set of all users.

\textit{Project} ($P$) is either a repository on GitHub or a page on Wikipedia.

\textit{Group} ($G$) is a set of users $G \subset \mathbb{U}$  who have performed at least one \textit{contribution} in the same project during the observed time period, and $G \in \mathbb{G}$, where $\mathbb{G}$ is a set of all groups. Collaborators are the users who belong to a group, where $u_{i} \in G$ and $i=[1 \dotsc N]$. Note that a single user can be a member of multiple groups at the same time. We limit our research on groups with at most 20 collaborators on GitHub, as only 0.006\% of all groups have more than 20 collaborators. \an{To maintain the statistical significance, we ignore the large groups. Small number of groups with more than 20 collaborators combined with the substantial variability can yield statistically non significant results.}\footnote{The abundance of data allows a proper analysis of group sizes whose occurrence is relatively small, but is large enough in absolute numbers. For example, there are only 26 groups with exactly 20 members in our sample on GitHub.} A distribution of group sizes on Wikipedia is less skewed and we are able to analyze larger groups with up to 70 collaborators. The number of groups with more than 70 collaborators drops significantly with each additional collaborator.

\textit{Group size} ($N$) is the number of collaborators belonging to a group as defined above, and it is simply the cardinality $N=|G|$.

\textit{Contribution} is either \textit{push} in GitHub or \textit{edit} in Wikipedia

\textit{Work} ($w$) quantifies the contributions of a user. The work of a group is the sum of all contributions of its members in the same project, $W= \sum_{i} w_{i}$.

\textit{Effective group size} ($n$) accounts for the relative number of contributions of group members in order to correct for the possible skewed distribution of work. It is defined as $n=2^{H}$, where $H= - \sum_{i=1}^{N}f_{i} \log_{2}f_{i}$, and $f_{i}=w_i/W$~\cite{Klug2016}.

\textit{Aggregate focus} ($F$) quantifies a fraction of contributions each member of a group brings to the project relative to their total contribution to all projects. It recognizes the fact that users usually work on multiple projects nearly simultaneously, allocating their time and effort among them. The contribution of a user to a project is a fraction of user's total contributions to all projects. To capture the \textit{aggregate focus} of a group we use $F = \sum_{i=1}^{N}c_{i}$, and $c_{i}=w_{i}/\sum_{i,k} w_{i,k}$, where $w_{i}$ is the work user $i$ performed on a project, and $w_{i,k}$ is the work user $i$ performed on a project $k$, where $k=[1,\dotsc \mathbb{T}]$. In other words, $c_i$ is a fraction of a total user's work dedicated to a particular project.

\section{Group performance and group size}
First,  we focus on the empirical analysis of the relation between group size $N$ (nominal number of collaborators) and group performance $\overline{W}$.

\begin{figure}
    \centering
    \includegraphics[height=5cm]{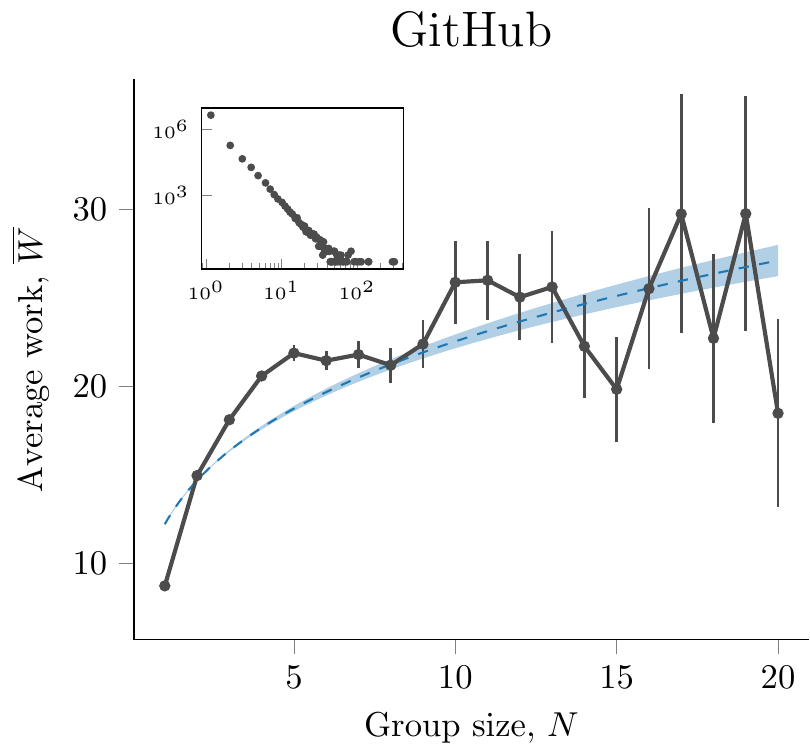}
    \includegraphics[clip,trim={0.7cm 0 0 0}, height=5cm]{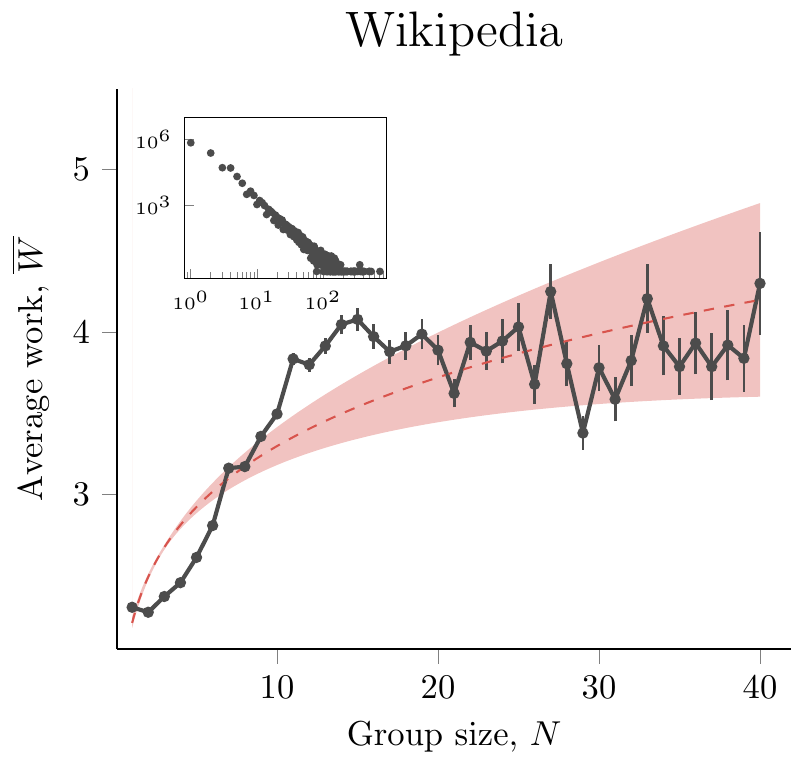}
    \caption{Average work per collaborator increases with group size. For larger groups, that trend fades out. Dotted line is a power-law fit, with an exponent $\alpha_{1}=0.28$ for GitHub and $\alpha_{2}=0.17$ for Wikipedia. It illustrates a super-linear trend suggesting that each additional collaborator increases the average group productivity. \textbf{Inset:} Distribution of group sizes has a long tail, where the groups of one (single contributors) make the overwhelming majority in both platforms.}
    \label{fig:work_vs_teamsize}
\end{figure}

Group performance is calculated as the average amount of work per member in a group as $\overline{W}=W/N$. Figure \ref{fig:work_vs_teamsize} shows the group performance $\overline{W}$ as a function of group size $N$. Each point on the plot represents the average work per collaborator for groups aggregated by their size, with vertical bars as $95\%$ confidence intervals. Dashed lines represent the power-law fit of the following functional form: $\overline{W}=N^{\alpha} \times c$, where $\alpha_{1}=0.28$, $\alpha_{2}=0.17$ and $c_{1}=e^{2.45}$, $c_{2}=e^{0.79}$ for GitHub and Wikipedia respectively. The positive value of the exponent $\alpha$ in both platforms, suggests super-linear scaling of group performance as number of collaborators grows. The total work $W$ as a function of group size can be obtained as
\begin{equation}
\begin{split}
W(N)    &   = c \int N^{\alpha} dN = \frac{1}{\alpha+1} N^{\alpha + 1} + C,
\end{split}
\label{eq:int}
\end{equation}
which yields
\begin{equation*}
\begin{split}
    W_{1}(N) & = 0.781N^{1.28}+C \text{\quad for GitHub, and} \\
    W_{2}(N) & = 0.854N^{1.17}+C \text{\quad for Wikipedia}
\end{split}
\end{equation*}

The scaling exponent we find is consistent with previous work showing that the median exponent is $4/3$ on open source software projects~\cite{Sornette2014}. To illustrate the importance of this improvement, consider a group of two collaborators on GitHub. A member of a group of size two makes on average 15 contributions, leading the group as a whole to produce 30 contributions on average. A  user who belongs to a group of size ten makes on average $\approx 22$ contributions, resulting in the group making about 220 contributions, on average. Although the number of collaborators grows only by a factor of five, their productivity increases eight-fold ($5^{1.28}\approx 7.85$), a super-linear increase in productivity. However, notice that such a scaling factor also indicates a decreasing marginal productivity of individual performance for increasing $N$.

One difficulty in interpreting these results is that the average work may not correctly reflect the distribution of work among collaborators. Instead, more work by larger groups could simply result from a few highly productive outliers skewing the average~\cite{Klug2016}. Another crucial aspect to consider is the possibility that this result is due to any confounding effect with productivity. For instance, the results may be distorted by selection bias~\cite{steiner2010importance}--larger, more popular projects could attract or recruit more experienced and active developers on GitHub or more specialized editors on Wikipedia. Thus, it is important to select covariates to control for selection bias. 

The next section is devoted to exhaustively test the robustness of the above results. First, we identify covariates that can help reduce any biases in the results. Second, in  Section~\ref{sec:ind_effort} we exploit the fact that users contribute to multiple projects in different group sizes to perform an analysis similar to a within-subjects design. The results are consistent: individuals in average intensify their work effort when groups get larger, up to the point when such a trend starts to decline.

\subsection*{Correcting for confounds}
We present a linear regression model of group performance to control for potential confounds among features related to the project and the group. We consider a set of intuitive features as the most obvious covariates for projects on GitHub and Wikipedia.

First, we perform a multivariate linear regression, using the Ordinary Least Square linear model. We take the average work of a group member $\overline{W}$ as a measure of group performance and a set of other available features as dependent variables. To correct for the difference in scale and estimate the relative contribution of all the variables, we scale the data so that all values fall into a range between zero and one. In order to achieve a statistically significant regression coefficient, we remove highly correlated predictors. The results of the OLS model (Table~\ref{tab:ols_group}) suggest that the group size, even when corrected for multiple confounding variables is positively related to the group performance in both platforms. Considering the fundamental differences between GitHub and Wikipedia, the features we chose are not the same for both platforms.\footnote{For example, a \textit{Fork} action on GitHub, which is good proxy for popularity, has no equivalent on Wikipedia.} 

\textbf{GitHub:} Group size $N$ is the most important factor explaining the group performance. Number of watchers and forks play the strongest role after group size. The number of watchers and forks are the indicators of repository popularity. More popular repositories attract more attention and subsequently more watchers and forks. With that in mind, it is reasonable to assume that more popular repositories also require more work, which is confirmed by the OLS model. Aggregate focus $F$ quantifies the fraction of work a user dedicates to a particular project compared to other projects. Larger average aggregate focus over all group members $\overline{F}$ indicates that a particular repository is one to which users focused mostly, hence yielding high performance of a group. Some factors, however, affect the work in a negative way, such as effective group size $n$. It suggests the existence of highly skewed work distribution among group members. It means that usually very small fraction of users do the majority of work. If the users who belong to a group work in many other projects at the same time, that will negatively affect the group performance. Here, it is quantified as the average number of projects, $G_s$.

\textbf{Wikipedia:} Similarly to GitHub, group size $N$ is an important predictor of group performance, while the average aggregate focus brings even more weight. Again, the number of projects users are working on will negatively affect the group performance.

\begin{table}
\caption{Ordinary Least Squares Linear Model for group performance. Outliers (above the 95th percentile in $\overline{W}$) were filtered out. *$p < 0.005$, $R^2_{1}=0.12$ and $R^2_{2}=0.1$ for GitHub and Wikipedia respectively}
\footnotesize
\begin{tabular}{lllllllll}
\toprule
\multicolumn{4}{c}{GitHub}& &\multicolumn{4}{c}{Wikipedia}\\ 
\cmidrule(r){1-4} \cmidrule(l){6-9}
\textit{variable}        & \textit{notation}              & $\beta_i$   & \textit{std.err.} & & \textit{variable}        & \textit{notation}              & $\beta_i$   & \textit{std.err.}     \\
\cmidrule(r){1-4} \cmidrule(l){6-9}
\textit{Intercept}            &             & 0.0315    & 0.000           & &\textit{Intercept}            &             & 0.1782    & 0.000              \\
\textit{Group size}& $N$                    & 0.9239    & 0.014          &  &\textit{Group size}& $N$                    & 0.3298    & 0.004              \\
\textit{Forks}& $F_o$                       & 0.2910    & 0.014          &   &\textit{Page age}& $A_p$                    & -0.1129   & 0.000            \\
\textit{Watchers}& $W_c$                    & 0.2680    & 0.013          &    &\textit{Avg. number of projects}& $\overline{G_s}$     & -0.1576   & 0.001           \\
\textit{Repo age}& $A_r$                    & 0.0249    & 0.001          & &\textit{Avg. aggregated focus}& $\overline{F}$         & 0.3975    & 0.009               \\
\textit{Effective group size}& $n$          & -1.0927   & 0.013           & && &             \\
\textit{Avg. number of projects}& $\overline{G_s}$     & -0.0591   & 0.003         & & & &              \\
\textit{Avg. aggregate focus}& $\overline{F}$         & 0.2011    & 0.006 & & & &                    \\
\bottomrule
\end{tabular}

\label{tab:ols_group}
\end{table}

\section{Individual work and group size}\label{sec:ind_effort}
After the analysis of group performance, we shift our focus to individual contributions. The amount of work among collaborators is not equally distributed. Therefore, higher performance of larger groups could stem from the highly productive individuals within the groups, who may be highly motivated to work on bigger projects. Also, larger projects might attract more active contributors: more experienced developers on GitHub or more specialized editors on Wikipedia. We address these questions by exploiting the fact that a significant fraction of users work on multiple projects simultaneously. By observing how users divide their effort across multiple projects enables us to measure the effect of group size on the allocation of effort. Specifically, given user $u$, who works on two projects $p_1$ and $p_2$, with different numbers of collaborators $N_1$ (for project $p_1$) and $N_2$ (for project $p_2$), where $N_2 = N_1 + 1$, and corresponding amounts of work $w_1$ for project $p_1$ and $w_2$ for $p_2$, what is the relative work gain $\beta = w_2-w_1$. 

The first approximation for solving this problem is again the OLS model, which will show how much the mean of the dependent variable $w$ changes given a one-unit shift in the independent variable $N$ while holding other variables in the model constant. The results of a simple multiple linear model which quantifies the individual work are given in Table~\ref{tab:ols_user} in Appendix~\ref{sec:simple_ols}.

One challenge in modeling how users allocate their effort across groups of different sizes is the significant individual variability among users: some are very productive regardless of group size, while  others make just few contributions. Also, users differ in number of projects they are contributing to at the same time. How can we compare the work gain of two very different users, where one works on tens of projects and another one in a handful of them? The second challenge is a potential non-linearity which OLS can not capture. Simple OLS model will give a single solution for all available values of independent variable, while there can be a significant difference in linear coefficients for different ranges of $N$. In the following sections we address these challenges. First we implement Mixed Linear Effect model (MLE) to correct for multiple differences between users. Secondly, we increase the resolution of the MLE to address the non-linearity in the model and preserve the level of explainability and statistical significance. Notice that the set of variables used to analyze the group dynamic is extended by the additional user-specific variables.

\subsection*{Mixed Linear Effect model}
Mixed Linear Effect (MLE) models are the extension of simple linear models to allow both fixed effects (to determine the impact of group size), and random effects (to allow for differences among individual users). Given that many users contribute within multiple groups of different sizes, we can perform an analysis equivalent to a within-subjects design, where one has repeated measurements of the subject at different treatment levels (group sizes)~\cite{lindstrom88}. By using a Mixed Linear Effect model we can account for the productivity variability existing across different contributors and focus on the average effect of group size for contributors working in multiple groups.

In order to estimate the general trends, we perform MLE analysis by binning users by different attributes: the number of projects they work in ($G_s$) and the level of their individual work ($w_j$) for GitHub and Wikipedia. Additionally we bin for number of followers they have ($F_l$) for GitHub only. Our analysis shows that regardless of the eg. number of projects, users tend to put more effort into projects with more collaborators. The results in Table \ref{tab:MLE_groups} show that the same applies to all the binning variables. The coefficients of group size obtained from the MLE models vary from 0.980 to 2.425 for GitHub and 0.005 to 0.014 for Wikipedia, indicating the positive relation between group size ($N$) and work ($w$). 

The abundance of data allows for even more fine-grained binning. To correct for the largest source of variability among users, we bin users at the individual level, treating each user as an independent bin. If a user works in multiple groups of different sizes, we calculate the linear coefficient as a relation between invested work and a group size. The users who work in multiple groups of equal sizes are excluded from the analysis. The obtained coefficient $\beta_1=2.786$ for GitHub and $\beta_1=0.066$ for Wikipedia is positive, suggesting that in general users prefer to invest more work in projects with more collaborators. Binning at the individual level is the ultimate binning strategy which eliminates many problems arising when trying to put subjects in a bin based on a single characteristic. This way we are able to estimate the trend regardless of the user's baseline work level, additionally correcting for number of confounds as shown in Eq.~\ref{eq:1} and Eq.~\ref{eq:2}. It is possible only when the number of observations is large enough to ensure the statistical significance. 

\begin{table}
\caption{Results of linear mixed effects model analysis for various bins of users. For all selected bins, the coefficient which quantifies the effect of group size $N$ is positive, suggesting that the marginal work gain for additional collaborator is positive regardless of the random variable, meaning that individual users in average put more effort in projects with more collaborators. Outliers (above the 95th percentile in $\overline{w}$) were filtered out. *$p < 0.005$}
\footnotesize
\begin{tabular}{lllllllll} \toprule
& \multicolumn{4}{c}{GitHub} & \multicolumn{4}{c}{Wikipedia}\\ 
\cmidrule(rl){1-1} \cmidrule(rl){2-5} \cmidrule(rl){6-9}
\emph{grouping} &\emph{\textbf{group size}}  & \textit{bins} & \textit{max bin} & \textit{mean bin} &\emph{\textbf{group size}}  & \textit{bins} & \textit{max bin} & \textit{mean bin} \\
\emph{by} &\emph{\textbf{coef.* $\beta_1$}} & & \textit{size} & \textit{size} &\emph{\textbf{coef.* $\beta_1$}} & & \textit{size} & \textit{size}      \\    
\cmidrule(rl){1-1} \cmidrule(rl){2-5} \cmidrule(rl){6-9}
No. of projects, $G_s$       & \textbf{2.412}     & 164   & \num{144464}    & \num{2387.8}    & \textbf{0.014}     & 1737  & \num{410047}    & \num{3702.2}    \\
Individual work, $w_j$          & \textbf{0.980}     & 5     & \num{221674}    & \num{78318.8}   & \textbf{0.005}     & 5     & \num{1.7}m    & \num{1.2}m       \\
Followers, $F_l$                & \textbf{2.425}     & 5     & \num{235303}    & \num{78318.8}   & & & &       \\ [0.05cm]
\cmidrule(rl){1-1} \cmidrule(rl){2-5} \cmidrule(rl){6-9}
User ID                         & \textbf{2.786}     & \num{318187}     & 535    & 1.2   &  \textbf{0.066}         & \num{198547}      & 1099    & 3.5         \\ 

\bottomrule
\end{tabular}

\label{tab:MLE_groups}
\end{table}

The marginal gain in individual work for each additional collaborator in a group is is much more significant on GitHub compared to Wikipedia. The platforms which we observe differ in many ways. Firstly, the entry ``cost'' is lower on Wikipedia as the amount of knowledge required to make an edit on a page is minimal. On the other hand, GitHub is mostly oriented to software engineers, and making a contribution to an existing repository requires a specific skill set. Secondly, the collaboration on GitHub requires a higher level of coordination. Editing a page on Wikipedia usually does not require an approval from the creator of the page, while on GitHub, the users who are allowed to make immediate change are authorized by the creator of the repository. The groups on GitHub are closer to \textit{structured teams} compared to Wikipedia. Those and many other differences could be used to explain the discrepancy in the marginal gain of invested work for each additional group member. 

\subsection{Chained LME model} \label{sec:chained_model}
The previous analysis implies there is a positive association between group size and work performed in a group. However, we are interested into finding out if a similar positive relation persists over the all ranges of group sizes. Therefore we increase the resolution of the linear model and observe the change in the linear trend. To do that, we perform the MLE analysis for various ranges of group sizes. One approach is to divide the independent variable $N$ to $k$ ranges and then to observe the linear trend within the ranges independently. We extend this principle by selecting overlapping ranges of $N$. We refer to it as the \textit{chained} linear model. Chained LME model makes up of multiple LME models built on incremental and overlapping ranges of an independent variable $N$. This method yields multiple coefficients for all values of $N$. That way, besides the prevailing linear trend we can also observe its change. How to chose the adequate width of ranges? Too narrow range would increase the resolution, but at the same time reduce the number of observations in each range. We chose the ranges to increase the resolution as much as possible and also to maintain a statistical significance at the same time. It turns out that for GitHub, a practical range width is 7 and for Wikipedia it is 10.

\begin{figure}
    \centering
    \includegraphics[width=0.9\linewidth]{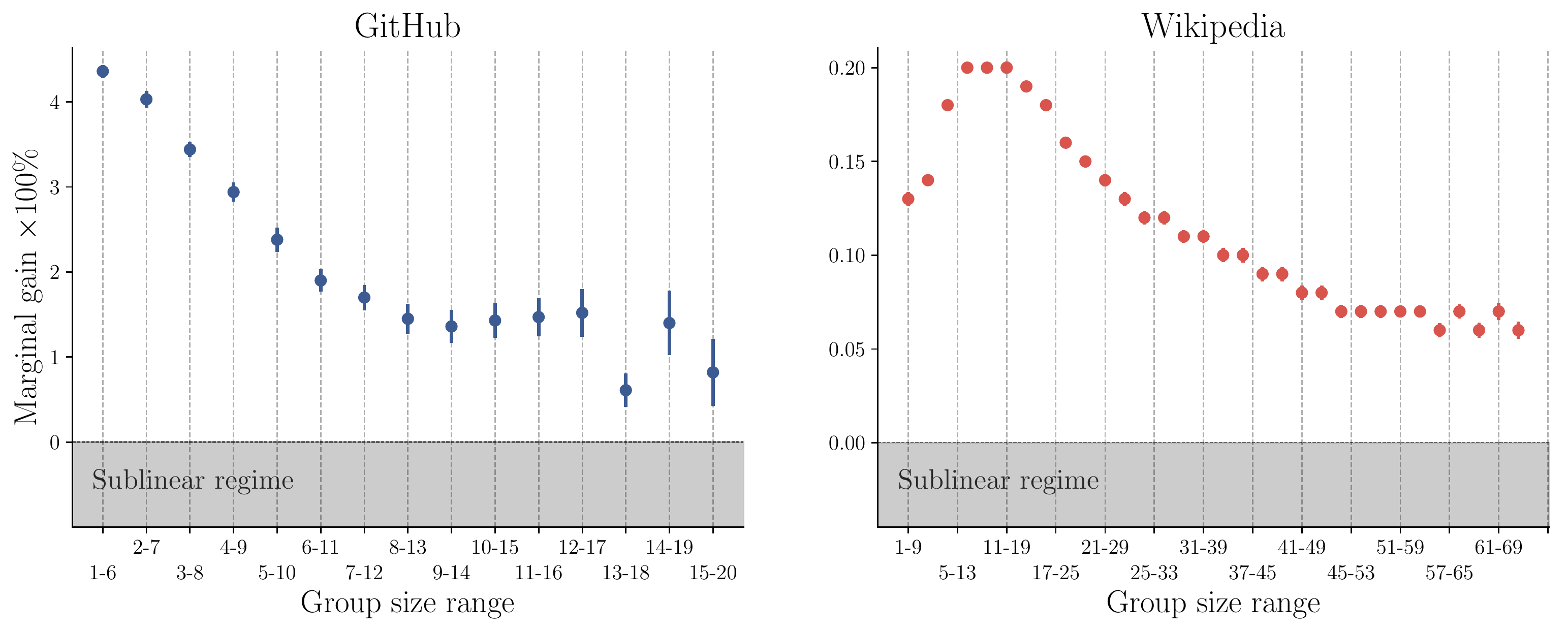}
    \caption{Marginal work gain for an additional collaborator. Each dot represents a marginal gain for additional collaborator ($y$-axis) depending on the range of the group size ($x$-axis). Vertical lines represent 95\% confidence intervals. The grey area on the bottom represents the sub-linear regime, where the work gain becomes negative.}
    \label{fig:marginal_gain}
\end{figure}

Again, we are primarily interested in a relationship between individual work within a group ($w$) and a group size ($N$). To correct for the most of the variability among users, each user is put in a separate bin and the data is filtered in a way to ensure that each bin consists of at least two records of different group sizes. That way we can estimate the trend for multiple group size ranges and aggregate it at the user level. Given the choice to work in groups of different sizes, the users increase the number of contributions in larger groups. We call the change in the amount of work for one unit change in group size a \textit{marginal gain}. The marginal gain in this case is a $\beta_1$ coefficient corresponding to group size $N$ variable in MLE model. It quantifies the relative change in the amount of work of a user for an additional collaborator.

To ensure the validity of the results we correct for the number of possible confounding variables\footnote{Please refer to Table~\ref{tab:notation} for definitions}. We use the Structured Sum-of-Squares Decomposition (S3D)~\cite{Fennell2018} algorithm to select the most important confounding variables. The procedure for selecting the most important factors is explained in Appendix~\ref{sec:factors}. For GitHub we chose: Effective group size ($n$), Watchers ($W_c$), User age ($A_u$), Aggregated focus ($F$), Owned repositories ($O_r$), Followers ($F_l$) and Number of projects ($G_s$). The analytical form of the MLE model used for each group size range on GitHub reads as:

\begin{equation}\label{eq:1}
\begin{aligned}
    w = {}    & \beta_1 N + \beta_2 n + \beta_3 W_c + \beta_4 A_u + \beta_5 F \\
              & + \beta_6 O_r + \beta_7 F_l + \beta_8 G_s + (1 | userID)
\end{aligned}
\end{equation}

For Wikipedia we chose: Effective group size ($n$), Number of projects ($G_s$), Total number of edits($W$), The largest group of a user ($N_{max}$), Page age ($A_p$), Average size of user's groups($\overline{N}$) and Number of created pages ($P_c$). The analytical form of the MLE model used for each group size range on Wikipedia reads as:

\begin{equation}\label{eq:2}
\begin{aligned}
    w = {}    & \beta_1 N + \beta_2 n + \beta_3 G_s + \beta_4 W + \beta_5 N_{max} \\
              & + \beta_6 A_p + \beta_7 \overline{N} + \beta_8 P_c + (1 | userID)
\end{aligned}
\end{equation}

\begin{table}
\caption{Results of the chained Mixed Linear Effect analysis. The obtained $\beta_1$ coefficient quantifies the relation between the group size $N$ and individual work $w$. MLE model has been applied on the selected ranges of group sizes. For all selected ranges, $\beta_1$ coefficient is positive, suggesting that users, when given choice, put more effort in projects with more collaborators. Such preference becomes less prominent for larger ranges. Outliers (above the 95th percentile in $\overline{w}$) were filtered out. *$p < 0.005$}
\footnotesize
\begin{tabular}{llllllllll} \toprule
\multicolumn{5}{c}{GitHub} & \multicolumn{5}{c}{Wikipedia}\\
\cmidrule(rl){1-5} \cmidrule(rl){6-10}
\emph{group size} &\emph{\textbf{group size}}  & \textit{observ.} & \textit{bins} & \textit{avg. bin} & \emph{group size} &\emph{\textbf{group size}}  & \textit{observ.} & \textit{bins} &  \textit{avg. bin}\\
\emph{range} &\emph{\textbf{coef. $\beta_1$}} & & & \textit{size}  & \emph{range} &\emph{\textbf{coef. $\beta_1$}} & & & \textit{size}       \\   
\cmidrule(rl){1-5} \cmidrule(rl){6-10}
1-6 & \textbf{4.36} & \num{173046} & \num{59880} & \num{2.9} & 1-9 & \textbf{0.13} & \num{176153} & \num{16436}  & \num{10.7}\\
2-7 & \textbf{4.03} & \num{106228} & \num{36593} & \num{2.9} & 4-12 & \textbf{0.17} & \num{172624} & \num{16917} & \num{10.2}\\
3-8 & \textbf{3.44} & \num{41754} & \num{15100} & \num{2.8} & 7-15 & \textbf{0.2} & \num{172406} & \num{18061}  & \num{9.5}\\
4-9 & \textbf{2.94} & \num{20590} & \num{7493}  & \num{2.7} & 10-18 & \textbf{0.21} & \num{172406} & \num{20297}  & \num{8.5}\\
5-10 & \textbf{2.38} & \num{11050} & \num{4093} & \num{2.7} & 13-21 & \textbf{0.19} & \num{172469} & \num{22017}  & \num{7.8}\\
6-11 & \textbf{1.9} & \num{6510} & \num{2455}  & \num{2.7} & 16-24 & \textbf{0.17} & \num{118087} & \num{17713}  & \num{6.7}\\
7-12 & \textbf{1.7} & \num{4370} & \num{1703}  & \num{2.6} & 19-27 & \textbf{0.15} & \num{80187} & \num{13827}  & \num{5.8}\\
8-13 & \textbf{1.45} & \num{2848} & \num{1139}  & \num{2.5} & 22-30 & \textbf{0.14} & \num{58696} & \num{11211}  & \num{5.2}\\
9-14 & \textbf{1.36} & \num{1953} & \num{797}  & \num{2.5} & 25-33 & \textbf{0.12} & \num{44558} & \num{9221}  & \num{4.8}\\
10-15 & \textbf{1.43} & \num{1287} & \num{552}  & \num{2.3} & 28-36 & \textbf{0.11} & \num{35204} & \num{7685}  & \num{4.6}\\
11-16 & \textbf{1.47} & \num{851} & \num{354}  & \num{2.4} & \multicolumn{5}{c}{\ldots}\\
12-17 & \textbf{1.52} & \num{588} & \num{244}  & \num{2.4} & 55-63 & \textbf{0.06} & \num{9195} & \num{2736}  & \num{3.4}\\
13-18 & \textbf{0.61} & \num{427} & \num{187}  & \num{2.3} & 56-64 & \textbf{0.07} & \num{8836} & \num{2640}  & \num{3.3}\\
14-19 & \textbf{1.4} & \num{385} & \num{165}  & \num{2.3} & 59-67 & \textbf{0.06} & \num{8133} & \num{2454}  & \num{3.3}\\
15-20 & \textbf{0.82} & \num{324} & \num{149}  & \num{2.2} & 62-70 & \textbf{0.07} & \num{6639} & \num{2069}  & \num{3.2}\\ 

\bottomrule
\hline
\end{tabular}
\label{tab:MLE}
\end{table}

The marginal gains for GitHub and Wikipedia are shown in Figure~\ref{fig:marginal_gain}, which reads as follows: on the $y$-axis there are various ranges of group sizes; on the $x$-axis there are the corresponding marginal gains for each range. For example, in case of GitHub, the user who works on a solo project is likely to contribute 4 times more if having one more collaborator. The results of LME analysis are also given in Table~\ref{tab:MLE}. For all selected group size ranges, the coefficient $\beta_1$ related to group size is positive, suggesting that users do invest more work in projects with more collaborators. However, the values of coefficients drop with increasing group size range. It suggests that the work per group member exhibits the increasing trend followed with a saturation. Notice that for the same relative change, there can be multiple marginal gains. The gain between groups of 5 and groups of 6 collaborators can have five different gains, depending on the range we observe. We address that by building a single model by averaging the gains obtained by multiple LMEs. \rev{We explore the robustness of these results to the choice of window size in Appendix \ref{sec:sensitivity_analysis}}

\subsection{Unified model}\label{sec:unified_model}
Multiple linear models can give an estimation of the effect of the group size to the individual work for various ranges of group sizes. Here, we attempt to unify multiple linear models into a single non-linear model. We use the results from the chained MLE model to give a characterization of the non-linear relationship between group size and productivity of individuals.

\begin{figure}
    \centering
    \includegraphics[height=5cm]{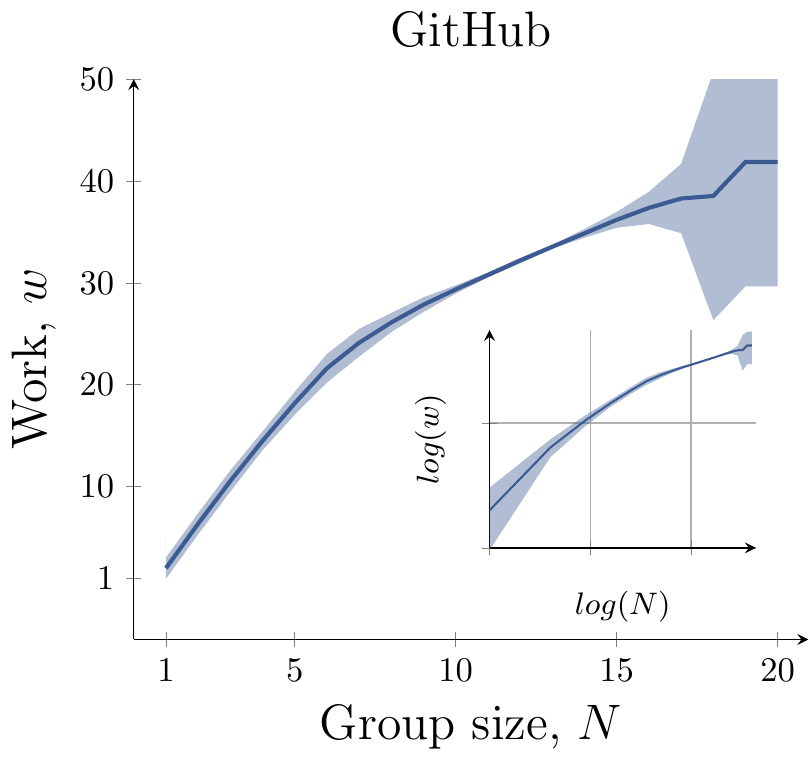} \quad
    \includegraphics[clip,trim={0.7cm 0 0 0}, height=5cm]{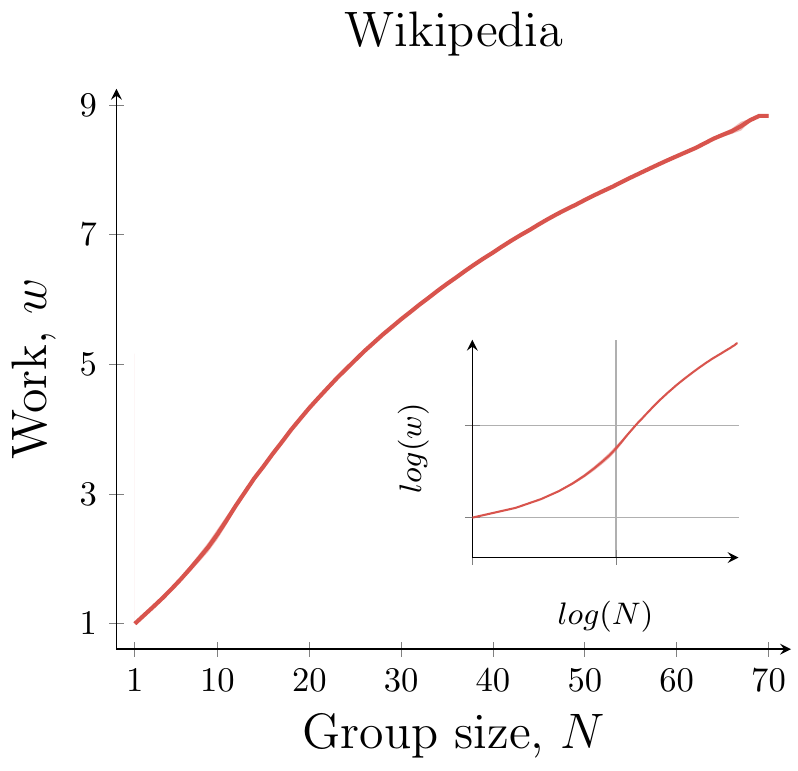}
    \caption{Model prediction of work as a function of team size. The means are estimated values obtained from the chained MLE model using the iterative process. The shaded area is 95\% confidence interval. Inset: The same data plotted on a log-log scale, suggesting that super-linear effect starts to fade out after $N \approx 10$}
    \label{fig:mlm_plot_coefficients}
\end{figure}

We define the ranges of group sizes as set of tuples $R=\{(1,1+v), (2, 2+v),\ldots,(m,m+v)\}$, where $v$ is a range width\footnote{$v=6$ for GitHub and $v=9$ for Wikipedia}, and $m$ is the number of ranges used for modeling. Each range $r_i$ has a lower bound $r_{i}^{l}$ which is the minimal group size in a range and upper bound $r_{i}^{u}$, the maximal group size in a range. For each range $r_i$, there is a corresponding mixed effects linear model $lm_i$ with a corresponding group size coefficient $\beta_i$. Group sizes are denoted as $k=[1 \ldots N]$. Obtaining a single model from a set of linear models is an iterative process which starts from the base value $wb_1$ and consecutive linear models are build on top of each other. It starts with $lm_1$ and a base value $wb_1=1$, which means we start building a non-linear model for a user who makes one unit of work in a solo team. Using $lm_1$ we then estimate the work for all group sizes in the range $r_1$. Then we move on to $lm_2$, and we chose the base value as the previously estimated value for group size 2. It is followed by estimating work for all group sizes in range $r_2$. In each iteration we chose the next linear model and the base value (starting point) for that model. The base value is chosen as a mean value of all previously estimated values for a starting group size. The iterative model is depicted in Figure~\ref{fig:model_iteration}. The work in a group of size $k$ estimated using $lm_i$ is:
\begin{equation}
    w_k|lm_i = f(wb_i,lm_i,k), \quad r_{i}^{l}<k<r_{i}^{u}
\end{equation}
The work in group of size $k$ is then:
\begin{equation}
    w_k = \overline{w_k|lm_i}, \quad \forall i \quad \text{where} \quad k \in [r_{i}^{l}\ldots r_{i}^{u}]
\end{equation}
The base work for a linear model $lm_i$ is: 
\begin{equation*}
    wb_i=\overline{w_k}, \quad k=r_{i}^{l}
\end{equation*}

The result after all the iterations is a set of estimated values of $w$ (individual work) for each group size. In Figure~\ref{fig:mlm_plot_coefficients}, we show the means and the 95\% confidence interval for the estimated work in various group sizes. 

The figure illustrates the functional form describing work $w$ invested by a member of a group as a function of the group size $N$. Consistent with our previous analyses we can observe that the marginal gain of having more collaborators is positive for all group sizes. It does however decrease rapidly to be negligible after certain point. Moreover, looking at the log-log plot in the inset of the figure, we have conclusive evidence that the super-linearity nature decreases beyond 10. Hence, rather than having a fix scaling exponent like in Equation~\ref{eq:int}, a better characterization of the data would be to have a decreasing scaling exponent.

\section{Conclusion}
Identifying the universal principles which apply to group productivity appears to be within the interest of various scientific disciplines. The complexity of human interactions and the heterogeneity of groups in terms of types and purposes makes this quest particularly challenging. Prior research demonstrates multiple conflicting results about the relationship between group size and productivity. In our attempt to untangle certain principles of collaborative work, we analyzed GitHub, a Web-based  collaboration platform for software projects and Wikipedia, online encyclopedia. First we identify some limitations of our work and then we summarize our main findings and contributions.

We acknowledge that our results on collaborative work should not be unconditionally generalized across domains. The data used in the study is collected from two different platforms: GitHub, used mostly by people who work in software development and Wikipedia which attracts users of many various interests. Each platform imposes its norms and dictates particular behaviors which can trigger domain-specific communication and work patterns. The groups in other work setups could exhibit a different behavior, hence one should be cautious about the absolute generalization of the provided results. \an{The definition of productivity varies among disciplines and authors. However, we equate the productivity with the amount of work, ignoring the nuances such as the time invested or the actual value of work.}

We used data covering more than 40 million actions performed by 2 million users working on 4.7 million projects on GitHub and 23 million edits performed by 700 thousand users on 2.6 million pages on Wikipedia. We explore the super-linear relationship of productivity as a function of number of collaborators on a project and show that adding a new collaborator in a group is often beneficial. An additional collaborator boosts the average work in a group, suggesting a synergistic nature in groups that causes users to perform significantly better than when working individually. 

We show that the productivity of a group as a whole is the result of increased productivity of all individual members, and not simply due to the presence of a few super-performing individuals. The heterogeneity and high variance of users' productivity makes it difficult to claim that the above phenomenon is more than an aggregate behavior and holds at the individual user level. However, by exploiting the fact the users work on multiple projects, we are able to demonstrate that people are more productive in larger groups. Given the choice to work on multiple projects, people put more effort into projects with more collaborators. The relation between group size and individual invested work is positive across all observed group sizes and it is particularly strong when the groups are still relatively small. For larger groups, that relation becomes less prominent and approaches zero. Overall, our results show that there is a super-linear relationship, but it is characterized by decreasing scaling exponent as groups become larger. This shows that the high productivity in larger groups is not the artifact of a small fraction of highly-productive individuals, but rather the result of a more universal preference pattern.

Besides the group size, we analyze a set of other factors which can drive user's productivity. We use S3D, a version of regression tree algorithm, to find the most informative features. Then, we extend this analysis by quantifying the linear relationships between the invested work and corresponding confounds.
%
\begin{acks}
The authors are grateful to the Defense Advanced Research Projects Agency  (DARPA), contract W911NF-17-C-0094, for their support. 
\end{acks}

%
\bibliographystyle{ACM-Reference-Format}
\bibliography{bibliography}

%
\newpage\appendix

\textbf{APPENDIX}

\section{Activity on projects}\label{sec:activity}
\begin{figure}[h]
    \includegraphics[width=0.85\linewidth]{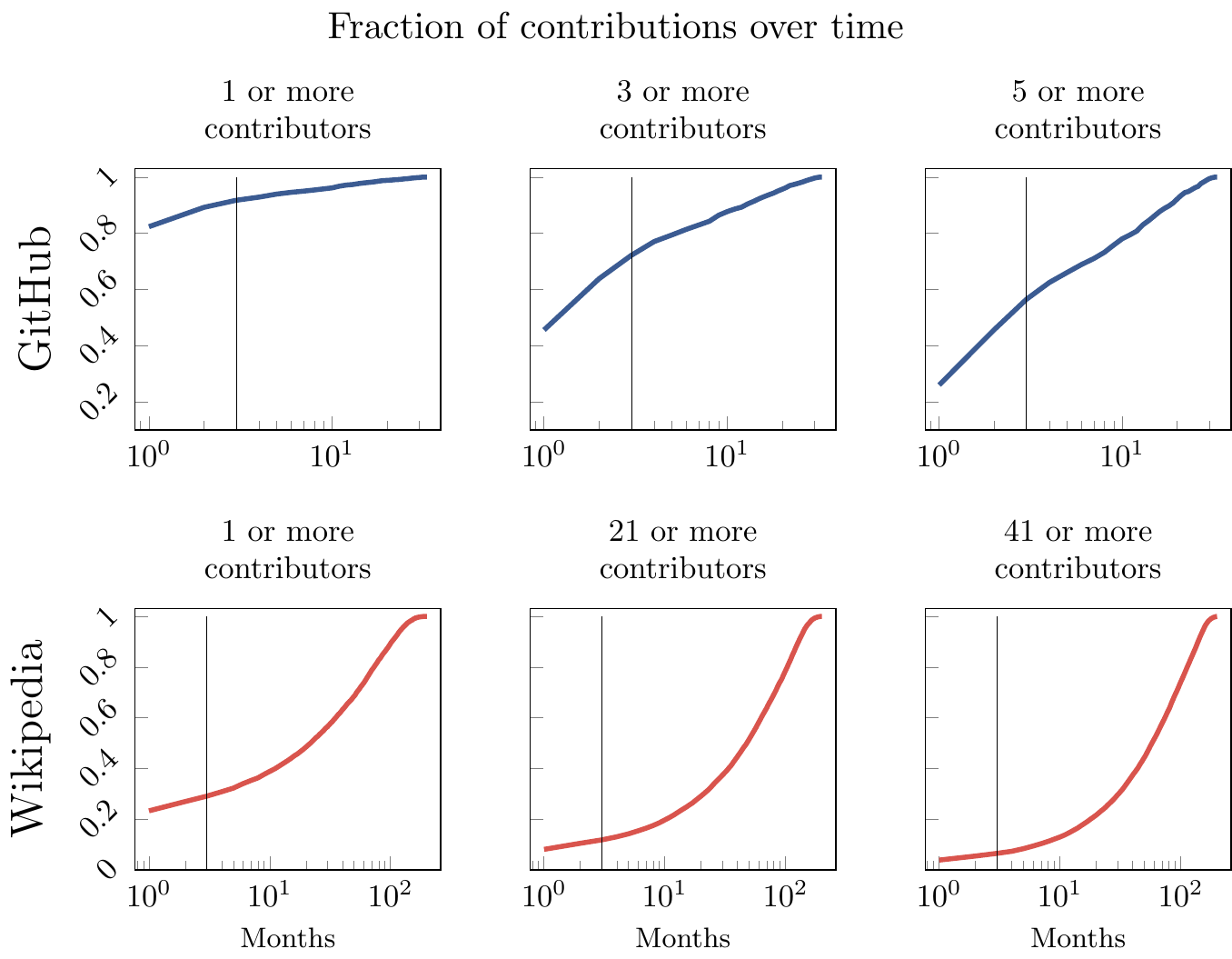}
    \caption{Contributions over time. Plots represent the cumulative fractions of contributions on Wikipedia and GitHub over time. Plots are divided based on the total number of collaborators on a project. The vertical lines mark the end of a 3 months period after creation of repository on GitHub or page on Wikipedia.}
    \label{fig:fraction_contributions}
\end{figure}

Pages on Wikipedia are relatively long-lasting. In average, Wikipedia article is 105 months old and the time difference between first and last edit is 93 months. Such a longevity allows contributions from multiple users over the period of many years. On GitHub, repositories get created with increasing pace making the average age of a repository only 3.5 months and with average difference between first and last recorded action only 2.6 months. Note that our data on GitHub does not contain any events recorded before Jan 2015, which affects some of those measures. Our strategy for selecting the relevant time frame for analysis considers several factors. The time frame we chose should: encompass a significant fraction of activities; reflect the reasonable time where a collaboration is more likely to occur (the time gap of the actions should not be too large); reflect the same phase of a project development; ensure the sufficient sample size. Since more than 90\% of all actions on GitHub repositories and more than 36\% of all edits on Wikipedia occur in the first three months (Figure~\ref{fig:fraction_contributions}), we choose the time frame of three months after the project creation.
\clearpage

\section{Sensitivity analysis of time window}\label{sec:sensitivity_analysis}
\begin{figure}[H]
    \centering
    \includegraphics[width=0.95\linewidth]{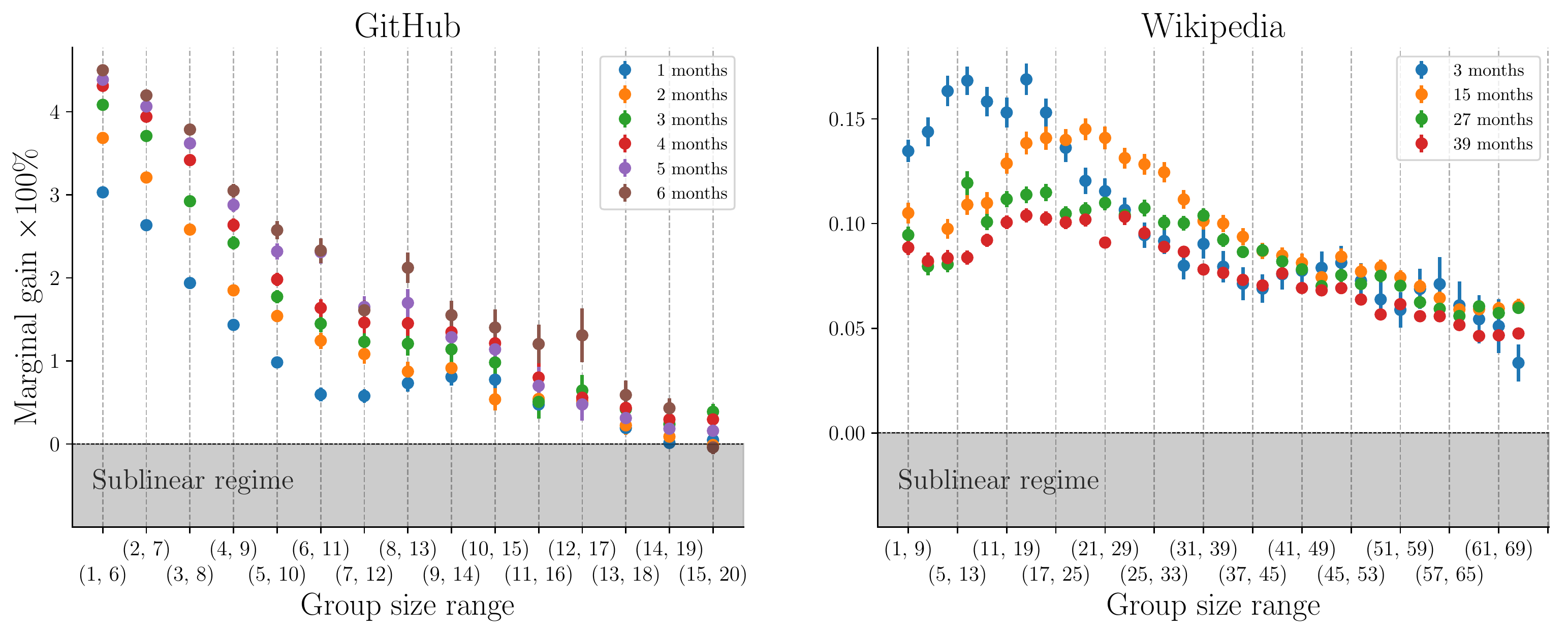}
    \caption{Marginal work gain for an additional collaborator for various time windows. Each dot represents a marginal gain for additional collaborator ($y$-axis) depending on the range of the group size ($x$-axis). Vertical lines represent 95\% confidence intervals. The grey area on the bottom represents the sub-linear regime, where the work gain becomes negative. Each color represents the marginal work gains calculated over different time windows.}
    \label{fig:marginal_multiple_gain}
\end{figure}

The most active period of a repository in GitHub is right after it's inception as the average activity drops over time. The same applies for pages in Wikipedia. In this paper, we focus most of our analysis on the first three months after the creation of a repository/page. However, the duration of the time window chosen after creation of repository/page can vary. In this section, we perform a sensitivity analysis on the duration $t$ of the chosen time window $T$. We let $t$ take values from 1 to 6 months for GitHub and from 1 to 44 months for Wikipedia, as the average age of Wikipedia articles allows such a wide range. For each value of $t \in \{1m \ldots 6m\}$ for GitHub and $t \in \{1m \ldots 44m\}$ for Wikipedia, we apply the Mixed Linear Effects Model to assess the individual work preference. First, we plot the marginal work gain for additional group member in Figure~\ref{fig:marginal_multiple_gain}. For the sake of clarity, we exclude some values of $t$ for Wikipedia. The visual inspection suggests that the analyzed phenomenon persists regardless of the time window chosen. 

\begin{figure}[H]
    \centering
    \includegraphics[height=5cm]{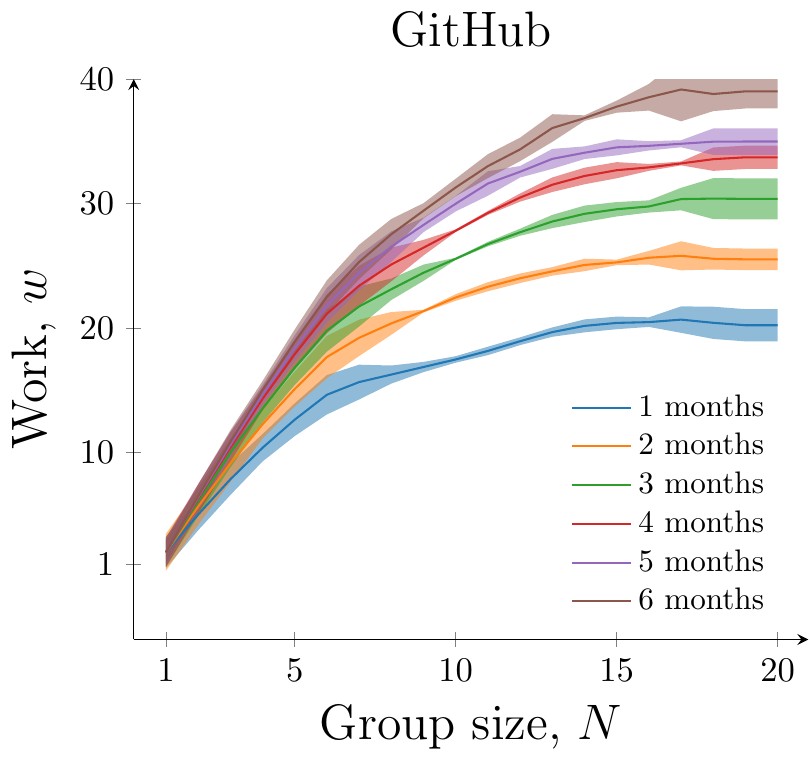} \quad
    \includegraphics[clip,trim={0.7cm 0 0 0}, height=5cm]{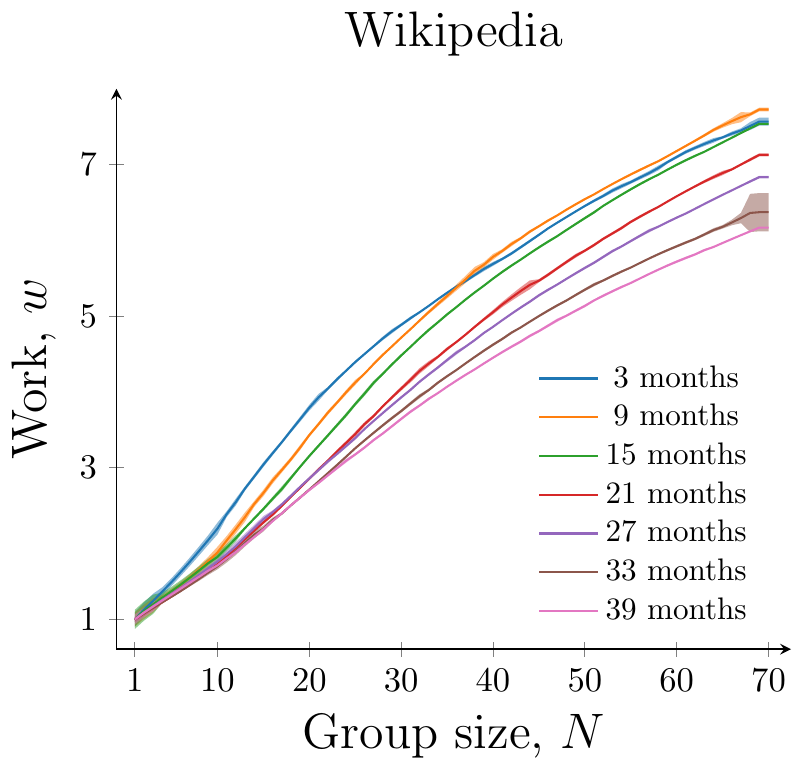}
    \caption{Unified models of work as a function of team size for various time windows. The means are estimated values obtained from the chained MLE model using the iterative process. The shaded area specifies the 95\% confidence interval. Each color represents the unified model calculated for different time windows.}
    \label{fig:mlm_plot_multiple_coefficients}
\end{figure}

Additionally, to better understand the effects of varying $t$, we build the unified model as explained in Section~\ref{sec:unified_model} for each value of $t$. The unified model allows to analyze the individual work invested in groups of various sizes. Multiple unified models with varying $t$ are depicted in Figure~\ref{fig:mlm_plot_multiple_coefficients}. We can represent the relation between two unified models as a function of time window $t$ as:

\begin{equation*}
    w(N,t_{i+1}) = \alpha w(N,t_{i})
\end{equation*}

\noindent We can approximate $\alpha$ by analyzing the resulting unified models and get $\alpha \approx 0.95$ for Wikipedia and $\alpha \approx 1.1$ for GitHub. Note that $\alpha$ is not a constant, but rather a function of $t$ which can be written as $\alpha(t_i) = \frac{t_i}{t_i+z}$ and $\lim_{t_i\to\infty} \alpha(t_i) = 1$. As the time window $t_i$ gets larger, the difference between two consecutive models $w(N,t_i)$ and $w(N,t_{i+1})$ gets smaller. 

\clearpage

\section{Factors of productivity}\label{sec:factors}

In order to better understand the elements which drive user's productivity, we focus on finding the most important factors which are able to explain the majority of the variability in the target variable $w$. To do that, we use Structured Sum-of-Squares Decomposition (S3D) algorithm~\cite{Fennell2018}. It is a variant of regression trees, that takes as input a set of features $\mathbb{X} = \{ X_i \}$, $i = 1 \dotsc M$ and an outcome variable $Y$ and identifies the optimal set of $m$ important features which are able to collectively best explain the outcome. S3D performs comparable to other tree-based methods in terms of prediction. However, its biggest advantage is its transparency and the ability to visualize and explain why predictions were made. 

The method partitions the values of each important feature such that $Y$ exhibits significant variations between blocks but small variations within each block. The procedure starts with the most informative feature and it's carried out recursively by splitting the next most informative feature, and so on. It stops when either of two criteria is met: a predefined number of features is reached or the increased variability of selecting the additional feature is not significant. Recursive data decomposition allows for the creation of the parsimonious model that explains not only the relations between the features and the target variable but the relations among the features themselves.  

The measure of variability used to explain the importance of features is $R^2$, which takes values between zero and one. Large values of $R^2$ indicate a larger proportion of $Y$ explained by the $X_i$. More informative variables have a larger $R^2$. During the iterative process, binning each variable adds to the total $R^2$ and the final $R^2$ is obtained as the proportion of the explained sum of squares to the total sum of squares:
\begin{equation}
    R^2 = \frac{\sum_{p \in P_{X_{j}}} N_p(\overline{y}_p - \overline{y})^2  }{\sum_{i=1}^N (\overline{y}_i-\overline{})^2}
\end{equation}

\begin{figure}
    \includegraphics[width=0.48\linewidth]{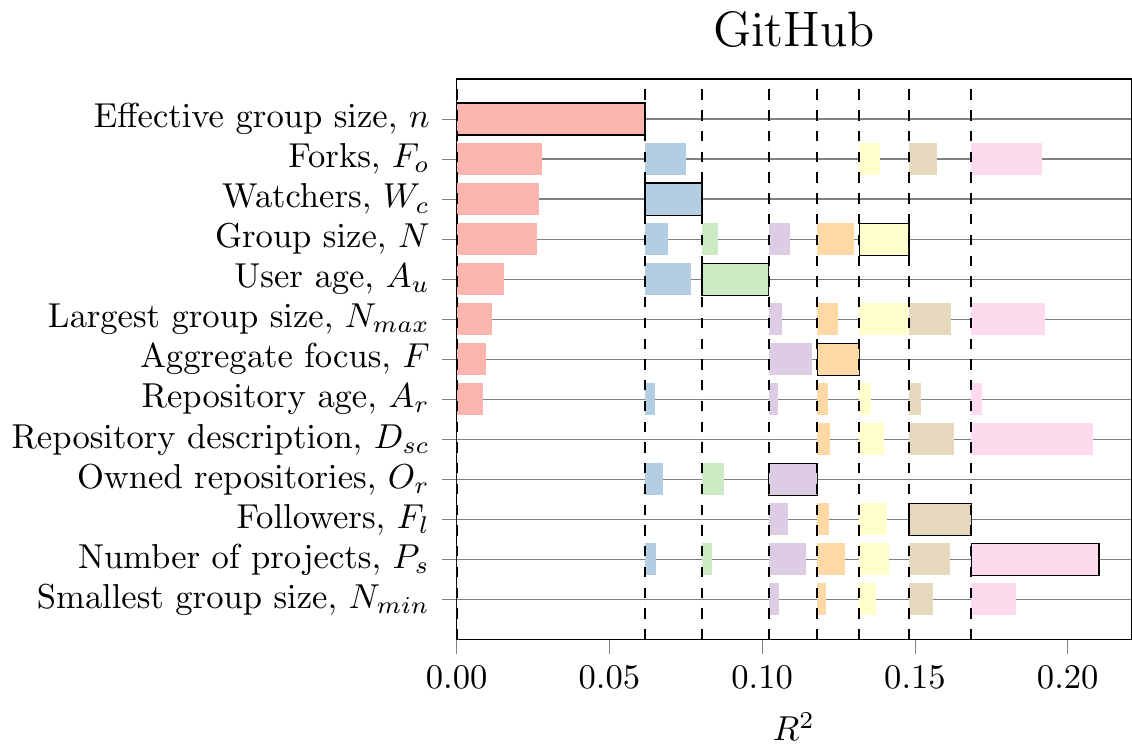} \quad
     \includegraphics[width=0.45\linewidth]{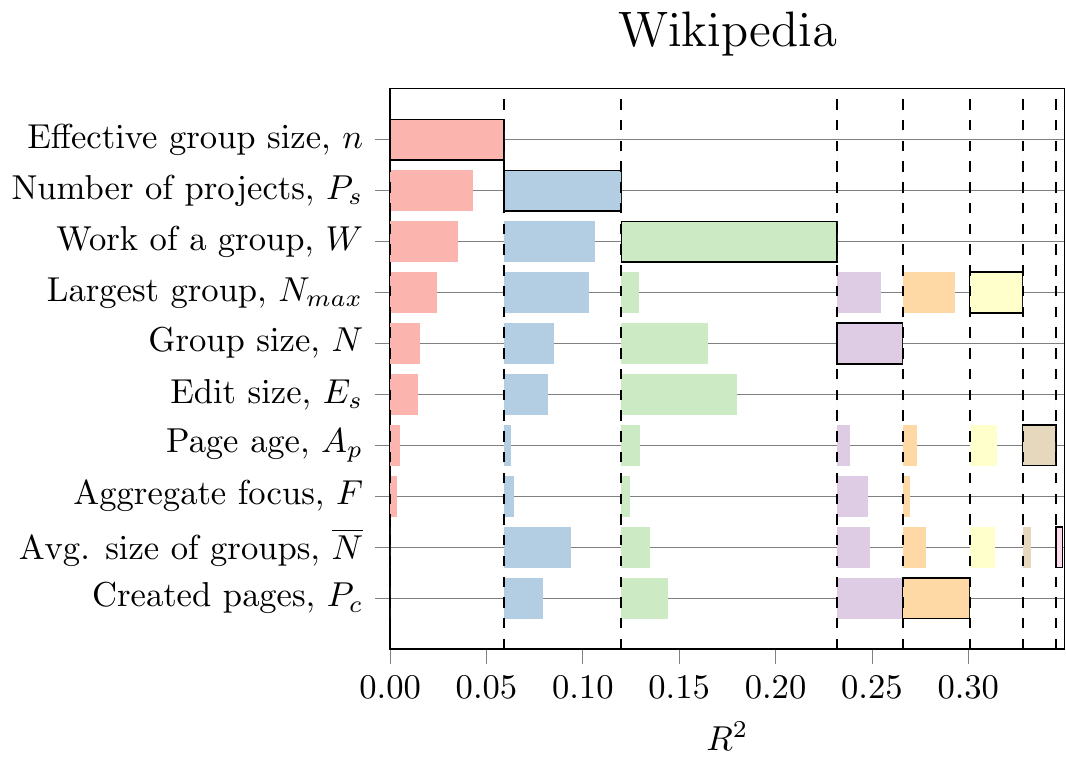}
    \caption{The amount of variation in the target variable $w$ explained by each feature in each iterative step of the S3D algorithm, measured by $R^2$. The feature that explains the most of the remaining variable at each step is marked with the black rectangle surrounding the associated bar. Notice that some features can have the same contribution at the same step of the algorithm, but are not being selected.}
    \label{fig:user_S3D}
\end{figure}

Very large values of $R^2$ do not always imply a very good predictive power of a model, as it can be the result of overfitting. In order to make sure that our model does not overfit, we performed k-fold cross-validation and optimize in-model hyper-parameter $\lambda$. That way we can be confident that we do not overestimate the importance of certain features. Here we focus on feature importance and use the model to learn which features can explain the target variable the best. In our case, the target variable is the amount of work of a user on a project, denoted as $w$. We explore one set of features for GitHub: \textit{effective team size}, \textit{watchers}, \textit{forks}, \textit{group size}, \textit{aggregate focus}, \textit{user age}, \textit{repository age}, \textit{number of projects}, \textit{largest group size}, \textit{smallest group size}, \textit{followers}, \textit{owned repositories} and \textit{repository description}; and some additional features for Wikipedia: \textit{work of a group}, \textit{edit size}, \textit{page age}, \textit{average size of groups} and \textit{created pages}. The results obtained by the S3D model are illustrated in Figure~\ref{fig:user_S3D}. The model was able to select eight most important features in eight runs for both platforms. Adding additional features would not significantly increase $R^2$.

Using S3D, we are able to identify the most important features and to build the tree-like model to estimate the work $w$ from the set of selected features. S3D quantifies the non-linear relationships among variables, by splitting them in the optimal number of bins. This approach, at the same time, makes it challenging to optimize the system for achieving the desired value of the target variable. For instance, \textit{effective team size} $n$ affects the work $w$, but it is unclear in which direction. The influence of $n$ to $w$ can be positive for one range of values of $n$ and negative for a different range. The model recognizes such relationships and applies binning accordingly. However, for the man-managed systems such as groups of people, keeping track on those subtle non-linear relations can become challenging.
\clearpage

\section{Multiple linear model for individual work}\label{sec:simple_ols}
To estimate how much the mean of the dependent variable $w$ changes given a one-unit shift in the independent variable $N$, while holding other variables in the model constant, we use the multiple linear regression and build an OLS model. The linear coefficients which correspond to all used variables are shown in Table~\ref{tab:ols_user}.

\begin{table}[H]
\caption{Ordinary Least Squares Linear Model on user's work. Outliers (above the 95th percentile in $w$) were filtered out.}
\footnotesize
\begin{tabular}{lllllllll}
\toprule
\multicolumn{4}{c}{GitHub} & & \multicolumn{4}{c}{Wikipedia}\\
\cmidrule(){1-4} \cmidrule(){6-9}
variable        & notation                                & $\beta_i$   & std.err. & & variable        & notation                                & $\beta_i$   & std.err.       \\
\cmidrule(){1-4} \cmidrule(){6-9}

\textit{Intercept}            &             & 8.0244    & 0.036     & &      \textit{Intercept}            &             & 3.6464    & 0.004                    \\ 
\textit{Effective group size}& $n$           & -2.0002   & 0.029    & &      \textit{Effective group size}& $n$          & -0.2526   & 0.001                \\
\textit{Watchers}& $W_c$                    & 0.0123    & 0.000     & &      \textit{Number of projects}& $G_s$          & -0.0002   & $4.74 \times 10^{-7}$               \\
\textit{Group size}& $N$                     & 1.4487    & 0.021    & &      \textit{Number of edits}& $W$             & 0.0001    & $3.67 \times 10^{-7}$                \\
\textit{Repo age}& $A_r$                    & 0.0056    & $6.5 \times 10^{-5}$  & &  \textit{Largest group}& $N_{max}$           & -0.0010   & $3.44 \times 10^{-7}$      \\
\textit{Aggregated focus}& $F$              & 0.4421    & 0.019     & &       \textit{Created pages}& $P_c$               & -0.0001   & $4.72 \times 10^{-7}$               \\
\textit{Number of projects}& $G_s$             & 0.0039    & $7.4 \times 10^{-5}$ & &  \textit{Page age}& $A_p$                    & -0.0542   & 0.000       \\
\textit{Followers}& $F_l$                   & 0.0002    & $3.7 \times 10^{-5}$ & &   \textit{Avg. size of groups}& $\overline{N}$         & 0.0091    & 0.000      \\
\textit{Owned repositories}& $O_r$          & -0.0005   & $8.1 \times 10^{-6}$ & &   \textit{Group size}& $N$                    & 0.1164    & 0.000       \\
\bottomrule
\end{tabular}

\label{tab:ols_user}
\end{table}

\clearpage
\section{Building unified model}
\begin{figure}[H]
    \includegraphics[width=0.95\linewidth]{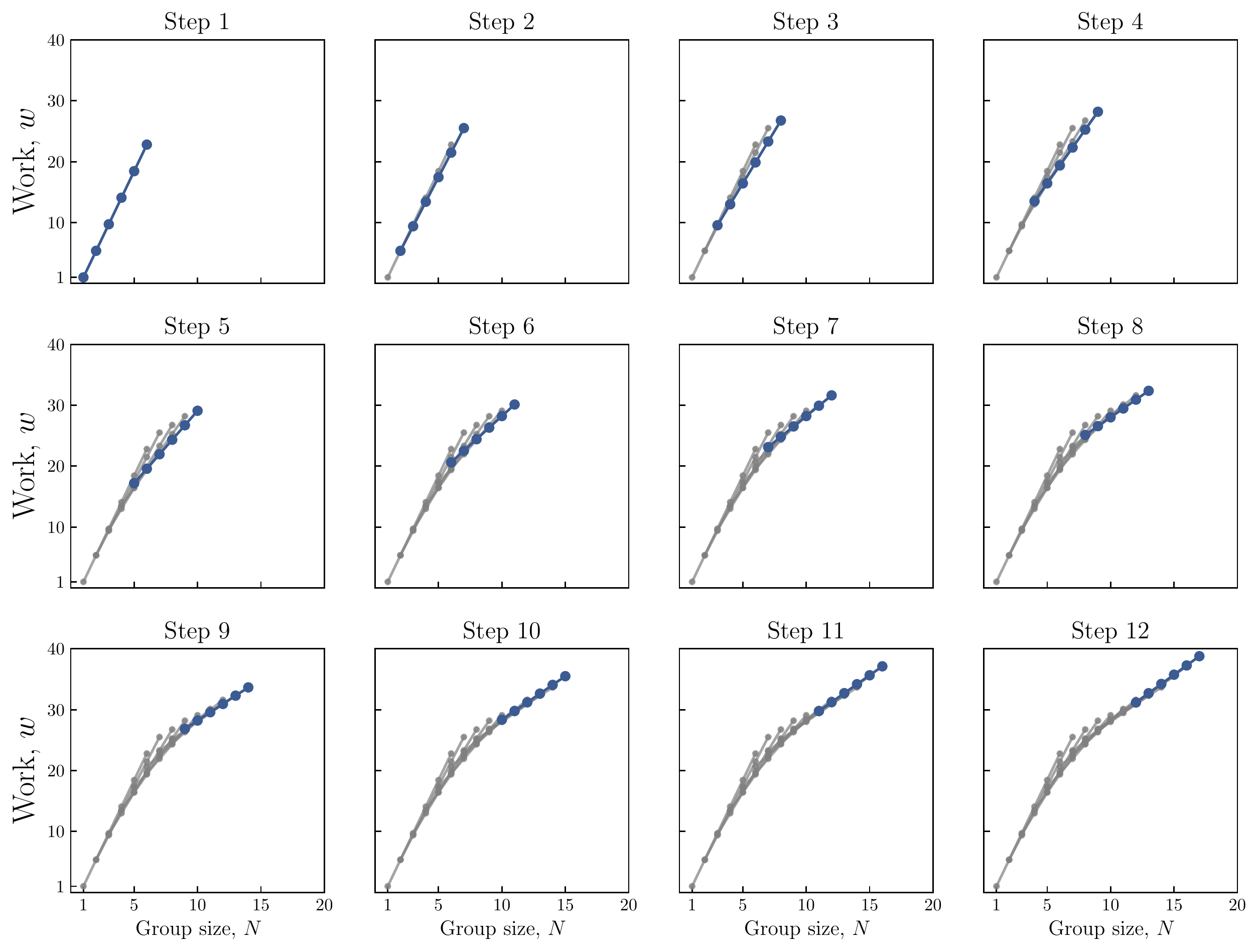}
    \caption{The iterations of building the unified model from the chained MLE model}
    \label{fig:model_iteration}
\end{figure}

Here we illustrate the iterative process of building a single model from the chained MLE model explained in Section~\ref{sec:unified_model}. Each subplot in Figure~\ref{fig:model_iteration} depicts a single iteration of the process. We start with base value $wb_1=1$, which means that we start building a non-linear model for a user who makes one unit of work in a solo team. Using $lm_1$ we then estimate the work for all group sizes in the range $r1$ (upper left subplot). We move on to $lm_2$, and we chose the base value as the mean of previously estimated values for group size~2. Then we estimate work for all group sizes in range $r2$ (upper row, second subplot from the left). In each iteration we chose the next linear model and the base value (starting point) for that model. The base value is always chosen as a mean value of all previously estimated values for a starting group size. The procedure is more formally defined in Section~\ref{sec:unified_model}.

\clearpage
\section{Scaling relation between alternative unit measures}\label{sec:wiki_scaling}
\begin{figure}[H]
    \includegraphics[width=0.55\linewidth]{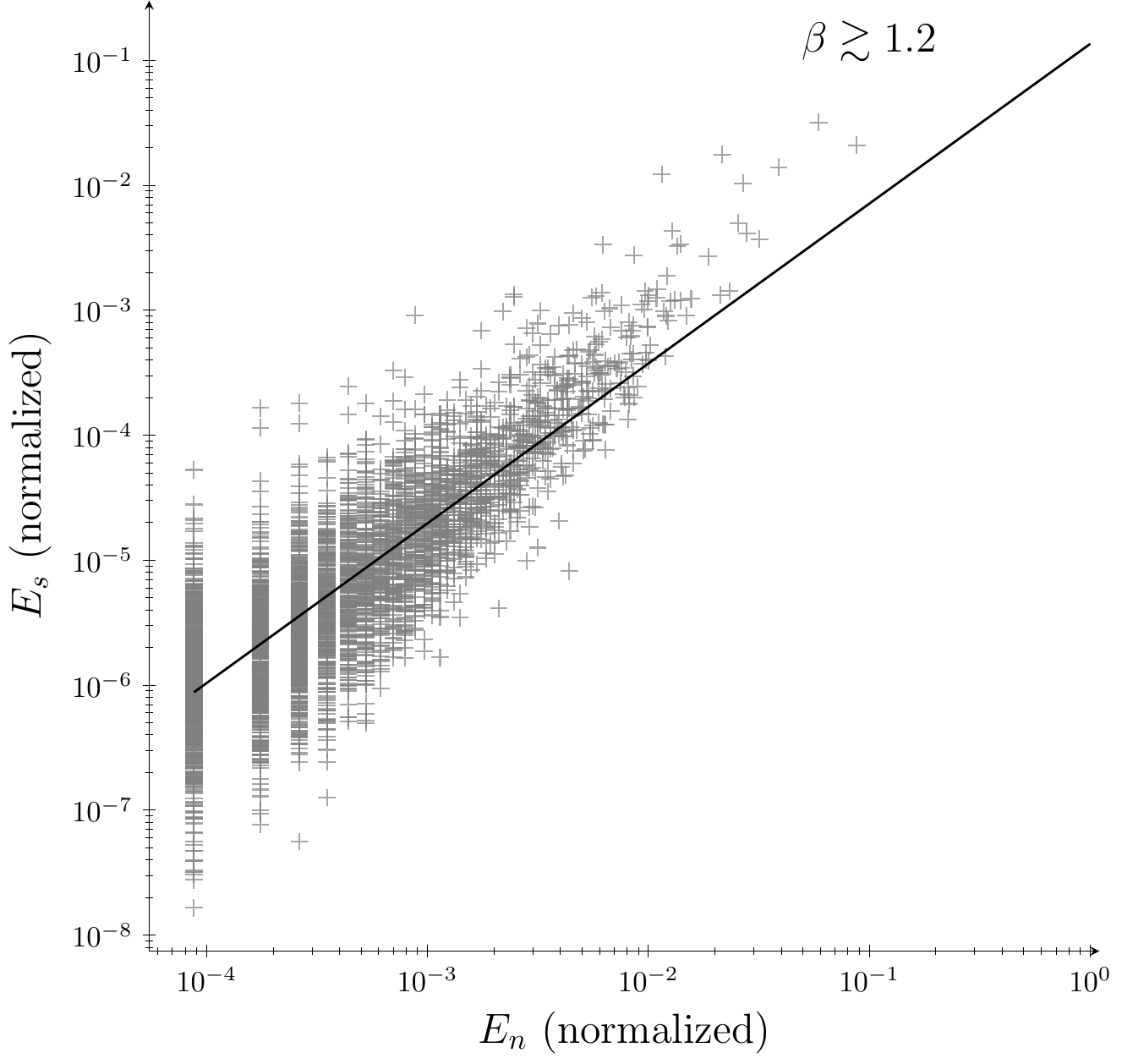}
    \caption{Scaling relation between edits and size of edits in Wikipedia. For Wikipedia, the scaling exponent is $\beta \gtrsim 1.2$ and $R^2 = 0.68$. The relation between number of edits $E_n$ and size of edits $E_s$ exhibits the similar scaling. It suggests that any of those measures could be used to measure the contribution.}
    \label{fig:wiki_scaling}
\end{figure}

\an{To measure the productivity of developers in GitHub, beside the number of \textit{push} actions we can use alternatives such as number of commits or lines of codes. Previous work~\cite{Sornette2014} has shown, on an open software development platform similar to GitHub, that the number of commits is highly correlated with lines of code. Since the vast majority of pushes consist of a single commit~\cite{Klug2016}, we stay on the safe side when using \textit{push} as the measure of productivity in GitHub. Similar applies for Wikipedia. An alternative unit measure to number of edits is the size of edit measured in bytes. Here, we test the relation between number of edits $E_n$ and size of edits measured in bytes $E_s$. Figure~\ref{fig:wiki_scaling} suggests slightly positive scaling relation $E_s = E_n^{\beta}$ with exponent $\beta \gtrsim 1.2$ and $R^2 = 0.68$. The results are similar to those reported in previous studies and show that as the number of edits grow, the size of edits does not decrease. This way we can stay confident in our choice to use the number of edits as the measure of productivity in Wikipedia.}

\end{document}